# Full-C-band, sub-GHz-resolution Nyquist-filtering (de)interleaver in photonic integrated circuit

Leimeng Zhuang[1*], Chen Zhu[1], Bill Corcoran[1,2], Maurizio Burla[3], Chris G. H. Roeloffzen[4], Arne Leinse[5], Jochen Schröder[6] and Arthur J. Lowery[1,2]

[1]*Electro-Photonics Laboratory, Electrical and Computer Systems Engineering, Monash University, Clayton, VIC3800, Australia*
[2]*Centre for Ultrahigh-bandwidth Devices for Optical Systems (CUDOS), Australia*
[3]*Institut National de la Recherche Scientifique (INRS-EMT), Montréal, Canada*
[4]*SATRAX BV, PO Box 456, Enschede, 7500 AL, The Netherlands*
[5]*LioniX BV, PO Box 456, Enschede, 7500 AL, The Netherlands*
[6]*School of Electronic and Computer Systems Engineering, RMIT University, Melbourne, Australia*
*\*leimeng.zhuang@monash.edu*

**Abstract**
Nyquist wavelength division (de)multiplexing (N-WDM) is a highly promising technique for next-generation high-speed elastic networks. In N-WDM, Nyquist filtering is an essential function that governs the channel spectral efficiency. However, most Nyquist filter implementations to date require either expensive, power-hungry digital electronics or complex arrangements of bulky optical components, hindering their adoption for important functions such as Nyquist channel shaping and reconfigurable optical add-drop multiplexers (ROADMs) for Nyquist super-channels. Here, we present a distinctive solution with low-cost, power-efficient, and simple-device natures, which is an on-chip optical Nyquist-filtering (de)interleaver featuring sub-GHz resolution and a near-rectangular passband with 8% transition band. This unprecedented performance is provided by a simple photonic integrated circuit comprising a two-ring-resonator-assisted Mach-Zehnder interferometer, which features high circuit compactness using high-index-contrast $Si_3N_4$ waveguide, while showing sub-picosecond optical delay accuracy enabling full C-band coverage with more than 160 effective free spectral ranges of 25 GHz across a bandwidth over 4 THz. Such devices show clear potential for chip-scale realization of N-WDM transceivers and ROADMs.

**Introduction**
The constant growth in demand for high data-rate telecom services, coupled with the need for an increasingly flexible and versatile network, poses an imminent need for a new generation of optical fiber networks. These networks must be not only able to support large capacity transmission, but also allow for flexible bandwidth resource management [1, 2]. In this critical fiber infrastructure, highly spectrally-efficient (de)multiplexing and optical channel control and routing with high frequency granularity are paramount for optimal operation. As such, Nyquist wavelength division multiplexing (N-WDM) is well placed to become a key multiplexing technology, as its well-defined rectangular channel spectrum intrinsically allows for add-drop operation by simple optical filtering [3-11].

A rectangular pass-band, high granularity optical filter is an important enabling technology for spectrally efficient fiber networks. In N-WDM, a sharp 'Nyquist' filter with a bandwidth equal to the channel baud rate is required, which is usually implemented by means of digital root-raised cosine (RRC) filters. Although digital implementations of Nyquist filtering are able to provide nearly perfect RRC filter responses with flexible roll-off factors [12], an optical Nyquist filter offers several advantages for practical applications. Optical Nyquist filters can be designed with multiple passbands to allow for multi-channel processing using a single device, promising a dramatically simplified system configuration – particularly when the filter provides wide frequency/wavelength coverage, e.g. over the full C-band as desired in applications. Moreover, optical filtering provides "line-rate" process capability without the need of bit-level access to the signals and avoids the use of high-speed digital electronics for digital signal processing and digital-to-analog conversion, which can considerably lower system latency, with potential savings in power consumption and cost. Perhaps most critically, optical Nyquist filters are an indispensable function in reconfigurable optical add-drop multiplexers (ROADMs) for N-WDM networks, the success of which is the key factor for the adoption of N-WDM in commercial optical network solutions.

Current optical Nyquist filtering approaches either provide too low resolution or too high complexity to satisfy the requirements of a flexible N-WDM network. Nyquist filtering methods involving nonlinear optical signal processes [13] have also been proposed, but require high complexity and/or power to operate over significant bandwidths. A more common approach, using spatial light modulator (SLM)-based optical filters, is typically limited by the filter spectral resolution in the order of 10 GHz [14-16], hindering the generation of Nyquist filters with sharp transition bands that match the common modulation bandwidths. Several schemes have been proposed to beat this limitation, by implementing optical signal processing functions on chip using photonic integrated circuits (PICs) [17-20]. Rudnick et al. demonstrated a modified SLM system featuring a significantly improved spectral resolution of 0.8 GHz [21], by using a dedicated arrayed waveguide grating as a dispersive element. Goh et al. reported an on-chip 8 × 1 optical Nyquist-filtering (de)multiplexer, based on a multi-tap delay-line circuit with electrically-controllable parameters [22]. Although both demonstrations show excellent functionality, the PICs used have a large number of delay lines, requiring extra tuning elements to achieve Nyquist filtering with sharp transition bands. Fundamentally, all the demonstrated circuit topologies to date are based on moving-average (MA) filters [23]. These filters use only feed-forward paths and require multiple delays with a maximum length one or two orders of magnitude times the circuit unit delay that is inversely proportional to filter bandwidth (e.g. 250 times in [21] and 56 times in [22]). The requirement for a large number of delays and tuning elements increases circuit complexity, loss, and power consumption, and renders the circuit more prone to optical phase and delay errors due to fabrication tolerance. This striking trade-off between the device yields and performance heavily hinders the practicality of such circuit topology for Nyquist-filtering purpose. One effective way to break this trade-off is to apply moving-average auto-regressive (ARMA) filters [23], where feed-back paths are used to generate an infinite number of delays without requiring a large number of delay lines in the circuits.

In this work, we present an on-chip optical Nyquist-filtering (de)interleaver, achieving high performance filtering with a significantly simplified circuit architecture. We change the MA-filter paradigm of Nyquist-filtering by shifting to an AMRA-filter solution that uses only three delay lines with a maximum delay of two times the circuit unit delay, featuring a two-orders-of-magnitude reduction of circuit complexity as compared to the circuit topologies in [21, 22]. The device consists of a two-ring-resonator-assisted Mach-Zehnder interferometer (2RAMZI) in high-index-contrast $Si_3N_4$ waveguide. Although discussions of this circuit topology have been performed in the context of conventional WDM and microwave photonic systems [24-26], it is challenging to realize such a device in high-index-contrast waveguide with low loss and precise control of circuit parameters, particularly when delay lengths in the order of centimeter are required to enable sub-GHz spectral resolutions. In fact, these properties are the key factors for the device to provide large stopband extinction and uniform performance over a large wavelength/frequency coverage [23]. Here, we show that our device is able to perform 12.5-GHz-passband Nyquist filtering with a sub-GHz resolution and a transition-band sharpness equivalent to a RRC roll-off factor of 0.08, and features sub-picosecond optical delay accuracy enabling full C-band coverage with more than 160 effective free spectral ranges of 25 GHz across a bandwidth over 4 THz. Further, we verify the Nyquist-filtering function of our device for N-WDM super-channel generation in transmission experiments, demonstrating negligible (0.1 dB) optical signal-to-noise-ratio (OSNR) penalty in comparison with digital Nyquist filter-generated signals. Moreover, pointing to a key network-level application, we show that our device can be used as a pre-filter before a commercial wavelength selective switch (WSS) to enable sub-GHz-resolution operations, demonstrating add and drop functions on N-WDM super-channels with < 1 dB OSNR penalty. These results show new possibilities for chip-scale realization of high-spectral-efficiency N-WDM transceivers and ROADMs that will play key roles in the next-generation high-speed elastic optical communication networks.

**Device description**

Figure 1a depicts a schematic of the Nyquist-filtering (de)interleaver circuit. It comprises a 2RAMZI, which is a 2 × 2 asymmetric Mach-Zehnder interferometer (MZI) with two ring resonators coupled to both of its arms. The ring resonators have an optical roundtrip path twice as long as the path difference between the MZI arms $\Delta L$, and the time delay for $\Delta L$ determines the free spectral range (FSR) of the circuit [23]. Mach-Zehnder (MZ) couplers are used throughout the circuit, allowing phase and coupling control with 7 chromium resistor-based heaters. Figure 1b and 1c show the circuit mask layout and a photograph of a demonstrator chip, respectively. The chip is fabricated in a commercial high-index-contrast $Si_3N_4$ waveguide technology platform (TriPleX [27, 28], see method). The waveguide circuit is designed with a FSR of 25 GHz and a chip area of $10 \times 4$ mm$^2$. The chip operates with end-face coupling, optimized for TE-polarization. Using fiber coupling, the total insertion loss of the chip measures to be about 9 dB (It is expected to decrease effectively when particular waveguide facet design or interposers are used to optimize the coupling).

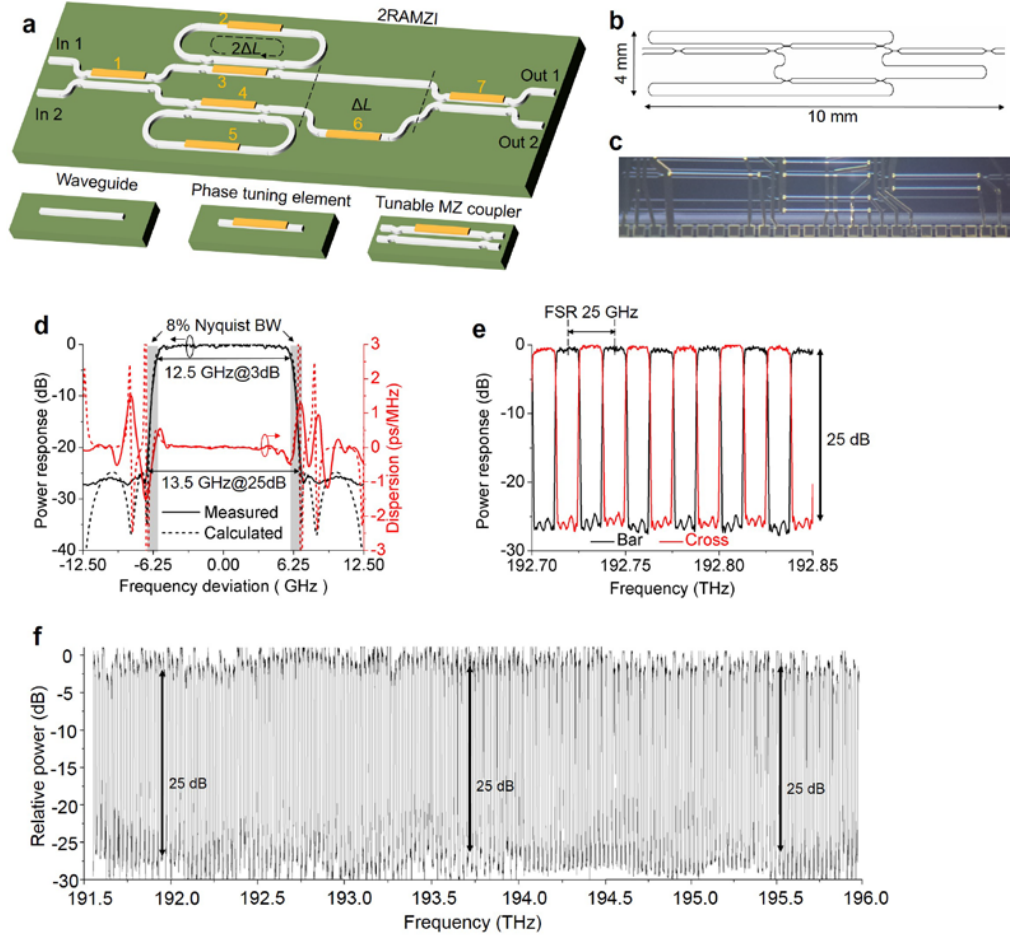

**Figure 1 | Nyquist-filtering (de)interleaver circuit and characterization. a,** A schematic of the circuit, comprising a 2RAMZI and 7 phase tuning elements. **b,** Circuit mask layout of a proof-of-concept chip using $Si_3N_4$ waveguide. **c,** A photograph of a fabricated chip, showing the waveguide circuit (grey) and metal connections of the resistor-based heaters (brown). **d,** Measured passband power response and dispersion. **e,** Measured bar- and cross-port responses. **f,** Passband measurement over full C band.

Figure 1d depicts the measured passband power response and dispersion, showing a good agreement with theory (see method). The near-rectangular passband is characterized by a flat top, a -3-dB bandwidth of 12.5 GHz, and a -25-dB bandwidth of 13.5 GHz. In terms of Nyquist multiplexing, this means a -25-dB inter-channel isolation at 8% of the Nyquist frequency. Figure 1e depicts the measured bar- and cross-port power responses with excellent passband resemblance and frequency periodicity, which indicates good waveguide uniformity and low wavelength dependency as desired. Figure 1f depicts a passband measurement from 191.5 THz to 196 THz, showing the full C-band coverage of the device. The power fluctuation over the measured passbands is an artifact, resulting from a defect of the measurement method (see method). The device operates with an excellent passband-stopband extinction of 25 dB throughout a bandwidth over 4 THz, equivalent to more than 160 FSRs of 25 GHz. This means a remarkable optical delay accuracy of sub-picosecond achieved in high-index-contrast $Si_3N_4$ waveguides with lengths in the order of centimeter. This demonstration proves the viability of creating high-performance optical processing functions on chip in compact forms.

**Nyquist-filtering and multiplexing for N-WDM transmitters**
Using the demonstrator chip as a Nyquist-filtering interleaver, we experimentally demonstrated all-optical Nyquist filtering and multiplexing for N-WDM transmitters, based on sub-bands created using two different types of light sources, i.e. independent CW lasers and a phase-locked frequency comb. The frequency comb generation technique is analogous to the all-optical orthogonal frequency division multiplexing techniques exploited by various groups [29, 30], while the CW technique provides insight into the performance of our (de)interleaver in a system with conventional dense-wavelength-division-multiplexing (DWDM) transceivers. The two corresponding experiment setups are depicted in Fig. 2a and 2b respectively, both generating Nyquist-spacing super-channels comprising seven 12.5-Gbaud QPSK-modulated sub-bands, using two independent data streams for simplicity. As the (de)interleaver chip is optimized to operate in a single polarization, polarization management was performed throughout the setup to guarantee a uniform, single polarization state for all sub-bands.

Figure 2c and 2e depict the spectra of a single sub-band based on the two different light sources. For either CW laser- or frequency comb-based approach, the Sinc- or noise-like modulated spectra will both have significant portion falling in the neighboring passbands of the interleaver, meaning strong inter-sub-band crosstalk for super-channel generation. This issue can be avoided by pre-filtering the odd and even channels before combining them at the interleaver, e.g. using two conventional WDM multiplexers or Waveshapers to restrict the per-channel spectrum within the bandwidth of two times baud rate [31]. Comparing Fig. 2c and 2e, the frequency comb-based approach results in sub-bands with flattened tops, unlike the CW laser-based approach. As such, the comb-based approach provides a closer approximation to ideal Nyquist channels and is therefore expected to offer better system performance. Figure 2d and 2f depict the measured spectra of 7-sub-band Nyquist-spacing super-channels generated as illustrated in Fig. 2a and 2b, respectively. For the sake of simplicity, in the CW-laser case, we implemented the spectrum-restricting pre-filtering in the digital domain before modulation, while in the frequency-comb case, the pre-filtering was skipped in accordance with our demonstration of two-data-stream interleaving.

We performed back-to-back (B2B) transmission experiments to verify the performance of the signals in Fig. 2c to 2f. Figure 2g depicts the measured signal quality factor ($Q$) versus optical signal-to-noise ratio (OSNR) for the case of single sub-band transmission; Figure 2h for the case of 7-sub-band super-channel transmission, with the center sub-band being received. To compare with traditional electrical-domain Nyquist-filtering methods, we separately generated a CW laser-based super-channel with the sub-bands defined before modulation, using digital RRC filter with a roll-off factor of 0.08 and using electrical pre-emphasis to flatten the spectrum. For all signal generation schemes, the digital receiver used a single-polarization constant modulus algorithm to converge to matched filter solution [32], followed by differential Viterbi-Viterbi frequency offset estimation and Viterbi-Viterbi phase estimation algorithms to recover the signals [33]. In Fig. 2g and 2h, the measured $Q$ values of the optical approaches show good agreement with their digital counterparts, having an OSNR penalty difference of < 0.1 dB at $Q$ = 8.53 dB — error-free threshold for 7% forward-error-correction (7%FEC). This means that optical Nyquist-filtering using 2RAMZI is able to provide comparable performance as digital RRC filtering with a roll-off factor of 0.08. Figure 2g also manifests a 7%FEC-threshold OSNR penalty of 0.5 dB in reference to the unfiltered QPSK transmission. This penalty is due to the fact that the filtered sub-band introduces inter-symbol interference due to the passband dispersion as shown in Fig. 1d. In Fig. 2h, the two optical approaches show similar performance and the effect of the unflatten passband of CW laser-based approach is only noticeable for OSNR > 15 dB. Compared with the single sub-band case, the super-channels have an extra OSNR penalty of < 1 dB at the 7%FEC threshold, which indicates the effect of the crosstalk between the sub-bands. A further reduction of the crosstalk will require an interleaver passband shape with sharper transition band and stronger out-of-band suppression. A recent study has shown the possibility for achieving these performance improvements in such a device [25, 34]. The results above fully verify the Nyquist-filtering and multiplexing functions of the (de)interleaver chip as a viable replacement for the conventional digital implementations, and the successful generation of Nyquist-spacing super-channels shows the feasibility of our all-optical approaches for practical applications in N-WDM systems.

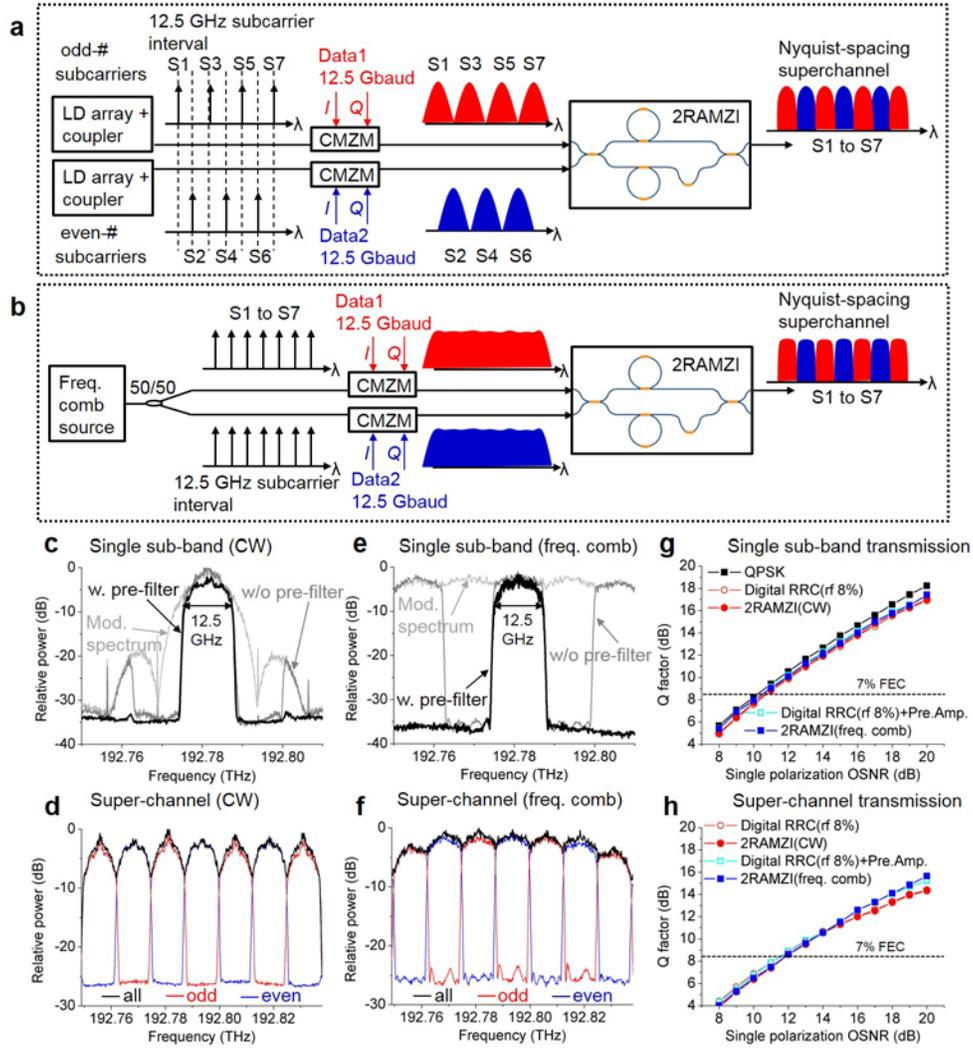

**Figure 2 | Nyquist-filtering and multiplexing for N-WDM transmitters. a** and **b,** Schematics of experiment setups. **c** and **e,** Spectra of single sub-band. **d** and **f,** Spectra of 7-sub-band Nyquist-spacing super-channels. **e** and **h,** Measured *Q* vs OSNR sweep.

**Sub-GHz-resolution ROADM for N-WDM super-channels**

Further, we demonstrate that by using our (de)interleaver chip as a fine-resolution pre-filter preceding a commercial WSS, an enhanced WSS (EWSS) with sub-GHz resolution can be created. This then enables add/drop multiplexing of sub-bands within an N-WDM super-channel. As illustrated in Fig. 3a, due to slow roll-off of the conventional WSS, dropping one sub-band from a super-channel causes severe inter-sub-band interference; however, with the deinterleaver preceding the WSS, the even and odd sub-bands will be first deinterleaved and can then be selectively dropped without such a problem.

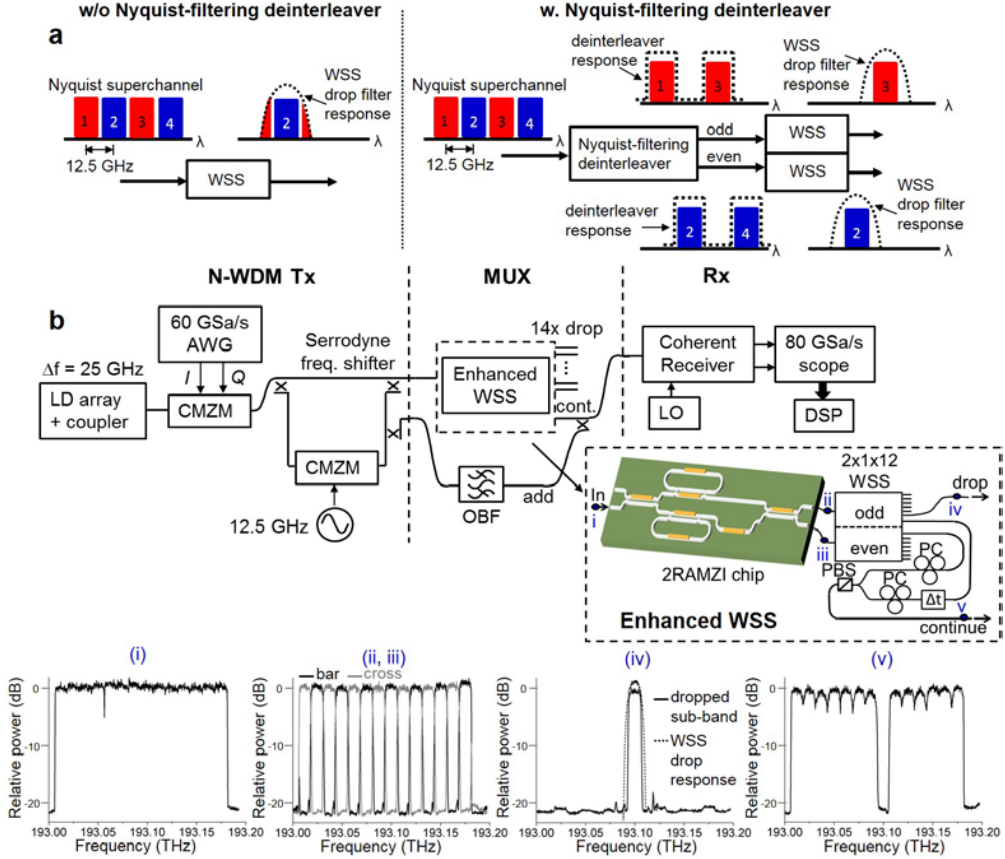

**Figure 3 | Deinterleaver chip-enhanced WSS. a,** Concept of sub-GHz-resolution N-WDM demultiplexing. **b,** Experiment setup for demonstrating key N-WDM ROADM functionality, including the EWSS (AWG: Arbitrary waveform generator, CMZM: complex I/Q modulator, OBF: optical band-pass filter, LO: local oscillator, DSP: digital signal processing). Insets (i-v) correspond to signal spectra at different points within the EWSS.

Figure 3b depicts the experiment setup. Here, the input optical signal to the EWSS is a CW laser-based Nyquist-spacing super-channel comprising 14 × 12.5-Gband sub-bands on a 12.5-GHz grid over a bandwidth of 175 GHz (with the leftmost sub-band, 'S1', at 193.0125 THz). The sub-bands are generated with a two-stream-interleaving pattern by means of the serrodyne frequency-shift method [35] using 7 CW lasers, where the Nyquist filtering is implemented using digital RRC filtering with a roll-off factor of 0.08 and electrical pre-emphasis in order to provide a sub-band spectrum shape similar to the optically generated one (see Fig. 2f). Figure 3b also shows the measured optical signal spectra at different positions of the EWSS, where the sub-band at 193.1 THz is dropped. The experiment also includes the emulation of 'add' operation, in order to fully demonstrate the viability of simultaneous, reconfigurable 'add' and 'drop' functions in a ROADM architecture using our EWSS. After the drop operation, the remaining odd and even sub-bands are de-correlated, polarization-aligned, and recombined into the 'continue' stream of the EWSS. A differential delay ≈ 10 ns is used for the de-correlation, which emulates a worst-case scenario that each sub-band experiences the maximum degradation from crosstalk. A replacement sub-band can then be added back with the 'continue' stream by means of passive coupling, where the added sub-band is obtained by filtering out a sub-band from a split of the serrodyne path and polarization-aligned with the 'continue' stream.

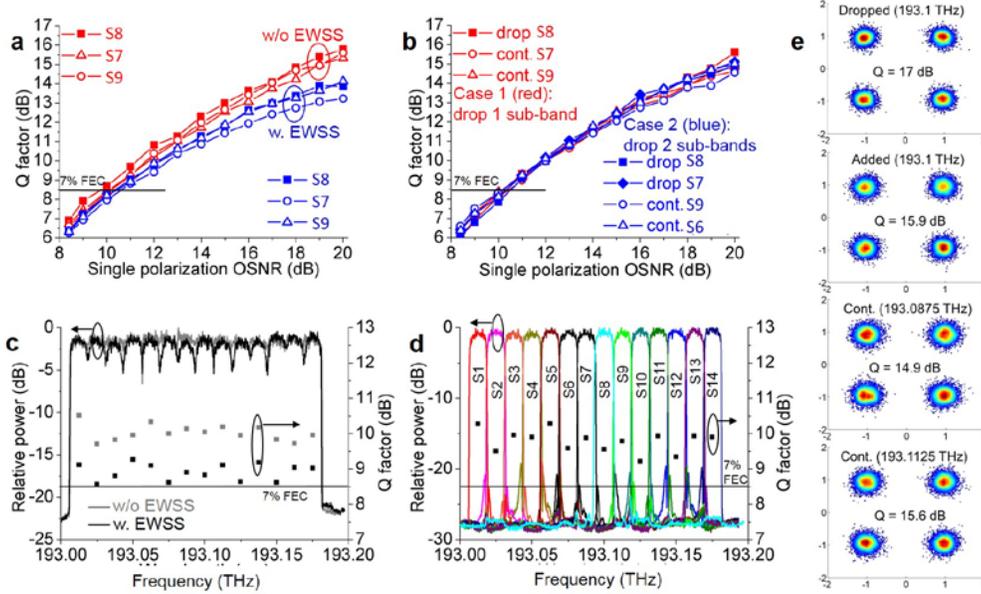

**Figure 4 | Reconfigurable add-drop multiplexing for N-WDM super-channels. a,** Measured $Q$ vs. OSNR for all sub-bands led to 'continue'. **b,** Measured $Q$ vs. OSNR for 'drop' and neighboring 'continue' (cont.) sub-bands with one or two sub-bands dropped. **c,** Spectra and $Q$ measurements for all sub-bands in 'continue' at a fixed OSNR of 11.5 dB. **d,** Spectra and $Q$ factor measurement for all sub-bands simultaneously dropped to separate EWSS drop ports at a fixed OSNR of 11.5 dB. **e,** Received constellations for 'drop', ''continue'' and passive 'add' sub-bands after multiplexing, without additional noise loading.

Figure 4a shows system performance with all sub-bands set to 'continue'. Signal $Q$ versus OSNR of three neighbouring sub-bands is plotted for two cases, i.e. with and without the EWSS included. Without the EWSS and noise loading, all three sub-bands show a $Q \approx 16$ dB, which is limited by the inter-sub-band crosstalk. When the EWSS is included, the three sub-bands show an OSNR penalty < 0.5 dB at the 7%FEC threshold compared to the case without EWSS. This penalty is incurred by the splitting and re-combining operation of the odd and even sub-bands. Figure 4b shows system performance when the EWSS is set to drop 1 or 2 sub-bands. The performance for the dropped ('S8' at 193.1 THz and 'S7' at 193.0875 THz) and neighbouring 'continue' sub-bands ('S9' at 193.1125 THz and 'S6' at 193.075 THz) is plotted against OSNR. The dropped and continued sub-bands have similar performance at the 7%FEC threshold, showing OSNR variation < 0.5 dB. Figure 4c shows the band-to-band performance comparison at a fixed OSNR of 11.5 dB. Here, similar $Q$s are observed, with a $Q$ variation of 0.7 dB for the case without the EWSS, and 1 dB for the case with the EWSS included and set to 'continue' for all sub-bands. Figure 4d demonstrates that the EWSS can demultiplex each of the 14 sub-bands to a drop port, where the band-to-band performance variation is only 0.7 dB for a fixed OSNR of 11.5 dB. Figures 4e exhibits the potential for simultaneous 'add' and 'drop' functions in the system. The constellations and extracted $Q$ values of the dropped, the passively added, and the nearest-neighbour 'continue' sub-bands are shown. In this result, there is no additional noise loading, and the *Q* of the 'continue' and passive 'add' sub-bands are within 0.5 dB of 15.4 dB, with the 'drop' sub-band at 17 dB. This shows that the 'continue' and passive 'add' sub-bands all experience similar cross-talk. The 'drop' sub-band is isolated from the super-channel before re-combination and so is not distorted by inter-channel interference, as reflected in its higher *Q*. The results in Fig. 4 show the feasibility of the desired 'drop' and 'add' functions using the proposed EWSS, and so prove the potential of the (de)interleaver chip for enabling key ROADM technologies for N-WDM super-channels.

**Discussion and conclusion**
The demonstrator (de)interleaver chip in this work operates with a WDM grid of 12.5 GHz. This can be changed in the circuit FSR design according to the bandwidth arrangement requirements of different applications. For the circuit design, additional ring resonators can be used in the RAMZI circuit to further improve the passband shape [25, 34]. However, the use of ring resonators means that the circuit introduces dispersion to the signal [23]. Previously, this has been considered as a road-block to the use of such filter designs in optical communication systems. However, with the advent of dispersion unmanaged transmission and digital dispersion compensation, this dispersion has little effect on the system performance, as observed in our experiments. Notably, the low circuit complexity of RAMZIs will benefit system construction greatly in terms of device yield, implementation of control and power efficiency compared to common MA filter designs. Like many other PIC applications, chip temperature stabilization is required during operation to guarantee the accuracy of circuit parameters. Depending on the used waveguide technologies, electro-optical tuning elements could be used to replace the thermo-optical ones, which will benefit the power consumption and compactness of the associated temperature control system. Further, polarization-diverse design could be considered in the future designs of RAMZI circuits to allow dual-polarization operations. Regarding photonic integration, chip-level integration of RAMZI circuits and electro-optics such as lasers, modulators and photodetectors will be highly promising for the development of N-WDM transceivers in terms of device compactness and packaging cost. For the realization of this, both monolithic integration and chip-to-chip micro-assembly technologies have shown clear potential in their current status of progress [36, 37].

In conclusion, we have reported a compact optical Nyquist-filtering (de)interleaver in high-index-contrast $Si_3N_4$ waveguide. The device features sub-GHz resolution, a near-rectangular passband with 8% transition band, and full C-band coverage with more than 160 effective free spectral ranges of 25 GHz across a bandwidth over 4 THz. In terms of system performance, the small penalties observed (0.1 dB in generation and < 1 dB in the EWSS) show that RAMZI devices provide a promising platform for optical filtering in N-WDM systems. Combining high-performance optical Nyquist-filtering and (de)interleaving functions, we anticipate that the reported device will open a new path towards compact, high-spectral-efficiency MUX and DEMUX devices for N-WDM transceiver and ROADM technologies that will play a key role in the next-generation, high-speed elastic networks.

**Methods**
**2RAMZI transfer function and configuration for (de)interleaver.** The transfer matrix of 2RAMZI, **H**, can be derived from a digital filter model in z-transform [23]:

$$\mathbf{H} = \begin{bmatrix} H_{11} & H_{12} \\ H_{21} & H_{22} \end{bmatrix} = \eta \begin{bmatrix} \sqrt{1-\kappa_R} & -j\sqrt{\kappa_R} \\ -j\sqrt{\kappa_R} & \sqrt{1-\kappa_R} \end{bmatrix} \begin{bmatrix} A_U(z) & 0 \\ 0 & A_L(z) \end{bmatrix} \begin{bmatrix} \sqrt{1-\kappa_L} & -j\sqrt{\kappa_L} \\ -j\sqrt{\kappa_L} & \sqrt{1-\kappa_L} \end{bmatrix} \quad (1)$$

with
$$A_U(z) = \frac{\sqrt{1-\kappa_1} - t^2 z^{-2} e^{-j\phi_1}}{1 - \sqrt{1-\kappa_1} t^2 z^{-2} e^{-j\phi_1}} \quad (2)$$

and
$$A_L(z) = \frac{\sqrt{1-\kappa_2} - t^2 z^{-2} e^{-j\phi_2}}{1 - \sqrt{1-\kappa_2} t^2 z^{-2} e^{-j\phi_2}} \cdot tz^{-1} e^{-j\phi_D} \quad (3)$$

where $z = \exp(-j\nu)$ with $\nu = [-\pi, \pi]$ representing the angular frequency normalized to the free spectral range (FSR) of the device ($\Delta f_{FSR} = 1/\Delta T = c/(\Delta L \cdot n_g)$ with $\Delta T$ the delay time for an optical path of $\Delta L$, $c$ the vacuum speed of light and $n_g$ group index of the waveguide); $t$ is the amplitude transmission coefficient for an optical path of $\Delta L$ (determined by the waveguide loss); $\eta$ is a complex coefficient which accounts for the general insertion loss and phase shift introduced by the MZI; parameters $\kappa$ and $\phi$ express the power coupling coefficient of the MZ coupler and phase shift, which are controlled via the heaters on our demonstrator chip. In the digital filter concept, the frequency responses between different input and output ports $H_{mn}$ ($m, n = [1, 2]$) correspondings to a 7$^{th}$-order infinite impulse response filter [23]. By properly setting $\kappa$ and $\phi$, the circuit synthesizes a (de)interleaver with the passband characteristics as shown in Fig. 2b. An example circuit parameter setting is shown below:

| Circuit parameter: | Definition: | Control heater: | Parameter value for (de)interleaver function: |
|---|---|---|---|
| $\kappa_L$ | Input coupler | #1 | 0.5 |
| $\kappa_R$ | Output coupler | #7 | 0.5 |
| $\kappa_1$ | Ring resonator coupling in the upper MZI arm | #3 | 0.87 |
| $\phi_1$ | Ring resonator phase in the upper MZI arm | #2 | $\pi$ |
| $\kappa_2$ | Ring resonator coupling in the lower MZI arm | #4 | 0.32 |
| $\phi_2$ | Ring resonator phase in the lower MZI arm | #5 | $\pi$ |
| $\phi_D$ | MZI lower arm phase | #6 | 0 |

**Waveguide technology.** The demonstrator chip described in this paper was fabricated using a $Si_3N_4/SiO_2$ waveguide technology (TriPleX™, proprietary to LioniX B.V., The Netherlands) in a CMOS-equipment-compatible process [27, 28]. This waveguide platform allows to modify the waveguide dispersion and polarization properties by changing the cross-section. The waveguides can be optimized to provide extremely low losses, around 0.01 dB/cm (which includes bending losses for a radius of 75 μm) at C-band wavelengths. Tapered facets can be provided to reduce fiber-chip coupling loss to lower than 1 dB/facet. Thermo-optical tuning elements were implemented using Chromium heaters and Gold leads, allowing for a tuning speed in the order of milliseconds.

**Experiments.** The frequency response of the photonic chip was characterized by means of wideband noise spectrum measurement. The wideband noise source uses the amplified spontaneous emission of erbium-doped fiber amplifier. The spectrum measurement was performed using a high-resolution spectrometer (Agilent 83453). Due to the lack of a noise source with sufficient bandwidth, the C-band response in Fig. 1f was obtained by means of measuring the responses of multiple 80-GHz-wide sections separately. For the measurement of each section, the input "white" spectrum was generated using 7 equal-spaced tunable CW lasers modulated at a symbol rate equal to their spacing. Due to the laser power fluctuation at different frequencies and wavelength dependency of the modulator, the measured passbands exhibit power fluctuations across the C band, however, this does not affect the passband-stopband extinction measurement. The chip was temperature-stabilized using a dedicated peltier element-based cooling system with maximum pumping power of 80W. Calibration and control of the tuning elements were conducted using multiple tunable voltage supplies (HAMEG HM7042-5). The CW lasers used in the experiments were narrow linewidth tunable lasers (AlnairLabs TLG-300M). The frequency comb was generated by overdriving a low-bias MZ modulator (Covega 20-GHz MZ modualtor) and optical filtering (Finisar WaveShaper). The data generation and digital pre-filtering were implemented using a 65GSa/s arbitrary waveform generator (Keysight M8195A). Complex 20-GHz bandwidth MZ modulators (Sumitomo DQPSK modulator) were used for QPSK modulation and serrodyne operation. A coherent receiver (u2t CPRV1220A) was used for signal reception. In the transmission experiments, pre-emphasis at the transmitter is neglected in favor of adaptive matched filtering at the receiver side. Before coherent detection, the optical signal is amplified and optically filtered (40 GHz bandwidth using Finisar WaveShaper), and noise-loaded with amplified spontaneous emission from erbium-doped fiber amplifier. The receiver output is digitized with a real-time, 80 GSa/s, 33-GHz bandwidth oscilloscope (Agilent DSO-X 928041), and offline processed. The digital signal processing flow resamples to 2 Sa/symb, applies spectrum-based frequency offset compensation, then matched filtering. A constant-modulus algorithm based adaptive equalizer and a Viterbi-Viterbi phase estimator are used before bit-error rate calculation.

**Acknowledgments**

This research work is financially supported by Australian Research Committee Laureate fellowship with grant no. FL13010041. The device under discussion is provided by LioniX B.V., SATRAX B. V. and University of Twente, The Netherlands.